%% file: main.tex
\documentclass[journal,10pt]{IEEEtran}
\usepackage[utf8]{inputenc}
\def\BibTeX{{\rm B\kern-.05em{\sc i\kern-.025em b}\kern-.08em
    T\kern-.1667em\lower.7ex\hbox{E}\kern-.125emX}}

\usepackage{graphicx}
\usepackage{color}
\usepackage{cite}
\usepackage{amsmath}
\usepackage{amssymb}
\usepackage{comment}
\usepackage{mathtools}
\usepackage{tabulary}
\usepackage{acronym}
\usepackage{upgreek}
\usepackage{algorithm}
\usepackage{algpseudocode}
\usepackage{placeins}
\usepackage[super]{nth}
\usepackage{dsfont}
\usepackage{subfigure}
\usepackage{tikz}
\usepackage{balance}

\usepackage{pgfplots}
  \pgfplotsset{compat=newest}
  \usetikzlibrary{plotmarks}
  \usetikzlibrary{arrows.meta}
  \usepgfplotslibrary{patchplots}
  \usepackage{grffile}
  \usepackage{amsmath}
\newtheorem{remark}{Remark}

\newcommand{\rev}[1]{{\color{black}#1}}
\newcommand{\dap}[1]{d^{n}(#1)}

\newcommand{\rmc}{\textrm{c}}
\newcommand{\be}{\begin{equation}}
\newcommand{\ee}{\end{equation}}

\allowdisplaybreaks
\input{acronyms}

\title{{Frugal Radio SLAM: Coherent Imaging with One Subcarrier and One Snapshot in Distributed MIMO}}
\author{\IEEEauthorblockN{
Yu Ge, Xin Tong, Nenad Vukmirović, Musa Furkan Keskin, Miljko Eri\'
c,  Petar M. Djuri\'c, Henk Wymeersch
}                                     % ...
\thanks{Yu~Ge is with the Research Laboratory of Electronics, Massachusetts Institute of Technology, Cambridge, MA, USA. (email: yuge@mit.edu)}
\thanks{Xin~Tong, Musa~Furkan~Keskin, and Henk~Wymeersch are with the Department of Electrical Engineering, Chalmers University of Technology, Gothenburg, Sweden. (emails: \{xinto, furkan, henkw\}@chalmers.se)}
\thanks{Nenad~~Vukmirović and Miljko~Eri\'
c are with the Department of Telecommunication, University of Belgrade, Belgrade, Serbia. (emails: \{nenad.vukmirovic, miljko.eric\}@etf.rs)}
\thanks{Petar~M.~Djuri\'c is with the Department of Electrical and Computer Engineering, Stony Brook University, Stony Brook, NY, USA. (email: petar.djuric@stonybrook.edu)}
\thanks{This work has been supported by the SNS JU project 6G-DISAC under the EU's Horizon Europe research and innovation Program under Grant Agreement No.~101139130, the Swedish Research Council (VR) through the project 6G-PERCEF under Grant 2024-04390, by the Ministry of Science, Technological Development, and Innovation of the Republic of Serbia under Grant No. 451-03-33/2026-03/200223, by the Palace of Science, Miodrag Kosti\'{c} Endowment, and by the Wallenberg Foundation and WASP Postdoctoral Scholarship  Program funded by Knut and Alice Wallenberg Foundation. OpenAI ChatGPT was used to polish and compress the text.}
}

\begin{document}

\bibliographystyle{IEEEtran}
\bstctlcite{IEEEexample:BSTcontrol}

\maketitle

\begin{abstract}
We study uplink radio \ac{SLAM} in a pre-calibrated phase-coherent distributed MIMO (D-MIMO) system at a minimal sensing-resource operating point: a single-antenna \ac{UE} transmits one narrowband pilot, and each %of $N$ 
single-antenna \acp{AP} provides one complex observation. We formulate coherent matched-filter imaging over a spatial grid, where the UE, virtual UEs associated with planar reflectors, and point scatterers share an equivalent-source representation. A greedy detect-and-cancel procedure separates these components. Simulations show that, at the considered noise level, weak-target recovery is limited primarily by cancellation residuals and model saturation, rather than by the noise floor.
\vskip0.5\baselineskip
\begin{IEEEkeywords}
Radio SLAM, D-MIMO, phase-coherent, narrowband, coherent imaging.
\end{IEEEkeywords}
\end{abstract}

\acresetall 

\section{Introduction}
In 5G and future 6G systems, localization and sensing are evolving from auxiliary features to fundamental capabilities \cite{procIEEE_2023_ISAC_6G}. Combined with the emergence of \acf{ISAC} as a unifying vision for future networks, this drives a paradigm shift from communication-centric radio infrastructure toward systems that can simultaneously localize users and sense objects in the environment \cite{procIEEE_2023_ISAC_6G,Fan_ISAC_6G_JSAC_2022}. Within this trend, a key ISAC approach is radio \acf{SLAM}, which jointly estimates the position of a \acf{UE} and builds a map of the environment \cite{Liza_SLAM_JSAC_2024,ge2025sensing}.

Conventional radio SLAM systems treat multipath not as interference to be suppressed, but as evidence of physical structures (walls, corners, and reflective surfaces) that can be localized alongside the UE \cite{multipath_SLAM_2016,massive_MIMO_SLAM_2019,Leitinger_SLAM_2019,ge2023mmwave}. From early ultra-wideband (UWB) demonstrations and progressing through millimeter-wave and 5G New Radio implementations, the field has matured considerably \cite{UWB_2013,massive_MIMO_SLAM_2019,Liza_SLAM_JSAC_2024}. The standard architecture employs a co-located multi-antenna \ac{BS} transmitting downlink signals to the UE, from which delay and angle measurements are extracted and fused to estimate the UE state and an environmental map on the UE side \cite{TWC_2019_Alessio,Liza_SLAM_JSAC_2024}. However, the performance of such systems is fundamentally limited by the aperture and geometry of a single BS, which restricts the angular resolution and the ability to resolve specular multipath from distant sources \cite{TWC_2019_Alessio}. Distributed multiple-input multiple-output (D-MIMO) systems, comprising spatially separated \acp{AP} that observe the same UE from geometrically diverse viewpoints, can offer a large effective aperture \cite{CF_MIMO_2017,6G_DMIMO_2025,tong_2026_imaging}. If their observations are phase coherent, they can be combined centrally for coherent imaging \cite{dario_imaging_2024_TWC}. This coherence is not free: it requires synchronization, calibration, and tracking of residual oscillator and hardware errors. 
Joint localization, synchronization, and mapping under phase-coherent distributed arrays is considered in \cite{JSTSP_Alessio_2025}, but relies on non-negligible bandwidth and multi-antenna \acp{AP}.

In this paper, we aim to understand how far phase-coherent D-MIMO imaging can be pushed when each \ac{AP} contributes a single antenna and a single narrowband snapshot, and which effects limit performance. This regime differs from wideband and/or multi-antenna phase-coherent localization \cite{vukmirovic2018position,vukmirovic2019direct,vukmirovic2021performance,JSTSP_Alessio_2025}, and from multipath-assisted SLAM built on delay/angle estimates \cite{multipath_SLAM_2016,Leitinger_SLAM_2019}, which do not apply when only one complex sample per AP is available. Our method is related to coherent matched-filter imaging, matching pursuit, and \ac{OMP} \cite{mallat1993matching,tropp2007signal}, as well as sparse source localization \cite{malioutov2005sparse}. We therefore do not claim a new general sparse-recovery algorithm; the novelty lies in this minimal-resource D-MIMO operating point and in identifying its limiting mechanisms.

Our contributions are:
%\begin{itemize}
    %\item
    (i) {we formulate a coherent imaging model for single-subcarrier, single-snapshot D-MIMO observations in which the UE, planar reflectors and point scatterers become equivalent sources sharing one steering-vector form, so that a single image tests all three hypothesis types;}
    (ii) {we specialize a greedy matched-filter / matching-pursuit-type detect-and-cancel procedure to this model, compare two cancellation estimators, and give the per-iteration complexity and a false-alarm-calibrated stopping rule that does not require the number of components;}
    (iii) {we derive closed-form expressions for the effects that limit this operating point: the cancellation residual and model saturation.}

\subsubsection*{Notations} {Scalars (e.g., $x$) are denoted by italics, vectors (e.g., $\boldsymbol{x}$) by bold lowercase letters, matrices (e.g., $\boldsymbol{X}$) by bold uppercase letters, and sets (e.g., $\mathcal{X}$) by calligraphic letters; the cardinality of $\mathcal{X}$ is $|\mathcal{X}|$. The transpose and Hermitian transpose are denoted by $(\cdot)^{\top}$ and $(\cdot)^{\mathrm H}$, respectively. The $\ell_2$ norm is $\|\cdot\|$, and the $i$th component of $\boldsymbol{x}$ is $[\boldsymbol{x}]_i$.}

\section{System Model}

\begin{figure}%[htbp]
\centerline{\includegraphics[width=1\linewidth]{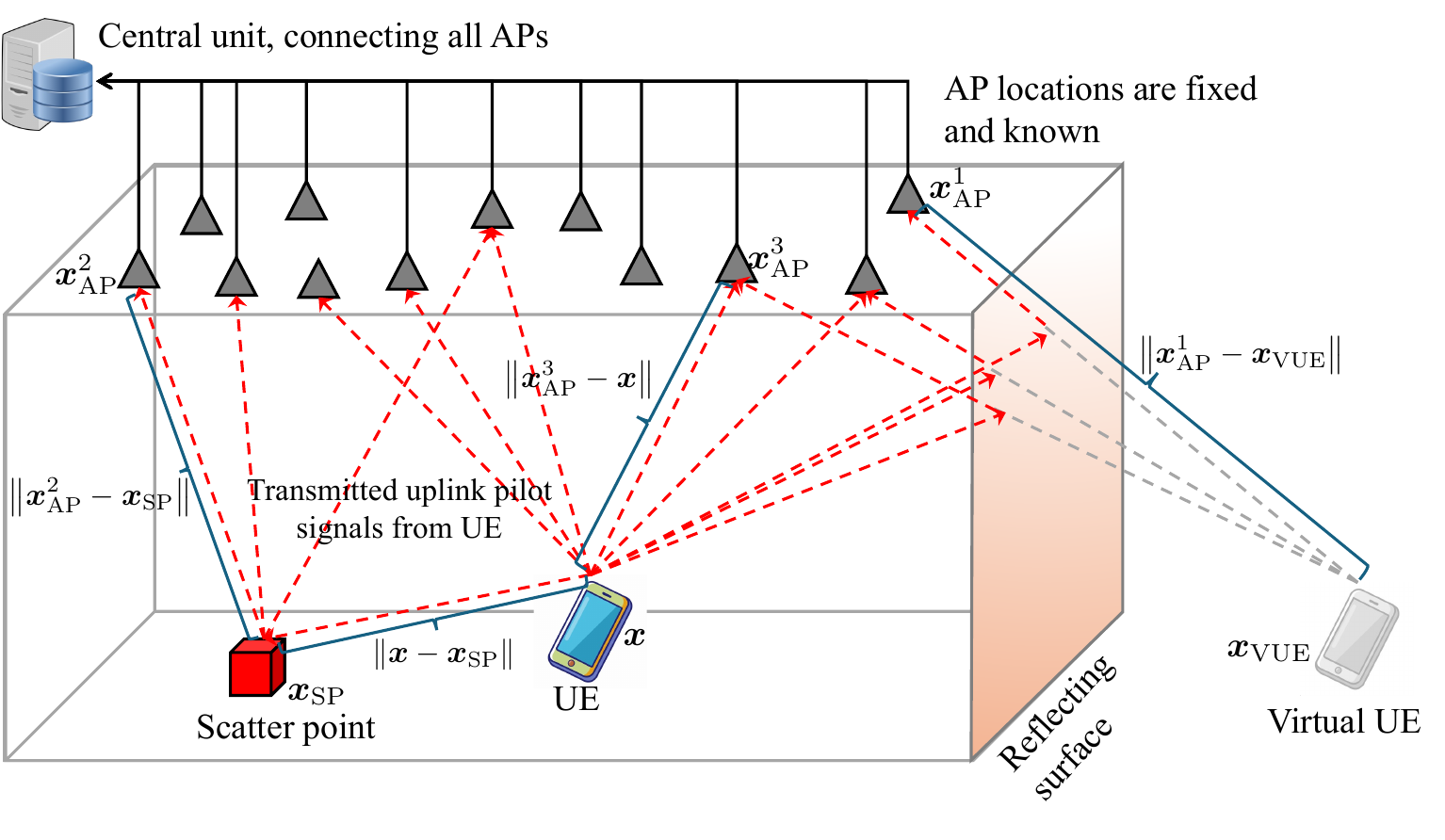}}
\caption{{Illustration of the uplink phase-coherent SLAM scenario with 12 APs, a UE, a reflecting surface, and a point scatterer. At a snapshot, the UE transmits an uplink pilot through the propagation environment, and the APs forward their observations to a central unit for phase-coherent processing, UE localization, and environment mapping.}
}
\label{fig:Scenario}
\end{figure}

\subsection{State models}
{We consider a static multi-\ac{AP}, single-\ac{UE} snapshot scenario, illustrated in Fig.~\ref{fig:Scenario}. The UE is in a known global coordinate system, with its position at $\boldsymbol{x}_{\mathrm{UE}}$. There are $N\gg1$ fixed single-antenna \acp{AP} at known positions $\boldsymbol{x}^{n}_{\mathrm{AP}}$. Perfect phase synchronization means that a calibration stage has already removed AP-dependent timing and carrier-phase errors over the snapshot. A central processing unit receives one complex pilot observation from every AP and performs the coherent combination. The model therefore trades low \emph{per-link sensing diversity} for network-wide calibration and fronthaul. }

{The static environment contains a finite landmark set $\mathcal{X}$. We include two idealized classes: a \ac{SP}, parameterized by a fixed position $\boldsymbol{x}_{\mathrm{SP}}$, and an infinitely extended planar specular reflector. The point model assumes isotropic scattering, while the reflector model retains only its single-bounce specular path. Thus, visibility, surface extent, polarization, diffuse clutter, and additional ceiling/floor/wall paths are excluded. Each reflector is represented, for the particular UE position, by a \ac{VUE}:}
\begin{align}
 {\boldsymbol{x}_{\mathrm{VUE}}=(\boldsymbol{I}-2\boldsymbol{\nu}\boldsymbol{\nu}^{\mathsf{T}}) \boldsymbol{x}_{\mathrm{UE}}
+2\boldsymbol{\nu} \boldsymbol{\nu}^{\mathsf{T}}\boldsymbol{\mu}.}
\end{align}
{Here $\boldsymbol{\mu}$ is any point on the surface and $\boldsymbol{\nu}$ is its unit normal. A VUE is a propagation-equivalent source for this UE/surface pair, not a complete estimate of surface extent or orientation from one snapshot. It is common to all APs for the given UE location, and the incidence point is the intersection between the AP--VUE line and the reflector.}

\subsection{Received signal models} \label{sec:receivedsignal}
{The UE sends a known pilot $\zeta$ on one retained narrowband subcarrier. The received sample at AP $n$ is the superposition of a direct \ac{LoS} component ($m=0$) and $M$ single-bounce \ac{NLoS} components ($m\in\{1,\ldots,M\}$):}
\begin{align}
    {y^{n} = \sqrt{E}\sum_{m=0}^{M} \rho^{m,n}e^{\jmath\theta^{m}}e^{-\jmath2\pi f_{\rmc}\tau^{m,n}}\zeta+w^{n}.}\label{eq:model1}
\end{align}
{Here $E$ is the transmission energy, $f_{\rmc}$ the carrier frequency, and $w^{n}\sim\mathcal{CN}(0,N_0)$ the receiver noise. The gain $\rho^{m,n}\geq0$ varies across APs. The AP-invariant phase $\theta^m$ contains the UE--network phase offset and, for NLoS paths, AP-invariant propagation and scattering/reflection phase. This is an \emph{isotropic, AP-invariant reflection/scattering-phase model}; it does not assume that physical scattering or reflection has zero phase shift. Residual AP-dependent phase errors are excluded by the post-calibration assumption.}

{Each component is compactly described by an \emph{equivalent source} position $\boldsymbol{x}^{m}$: the UE itself for the LoS path, the VUE for a specular reflector, and the SP for a point scatterer. With $\dap{\boldsymbol{x}}\triangleq\|\boldsymbol{x}-\boldsymbol{x}^{n}_{\mathrm{AP}}\|$, the delays and gains are listed in Table~\ref{tab:propagation}. Here $c$ is the speed of light, $\lambda$ the wavelength, $\Gamma\in[0,1]$ the reflection attenuation, and $\beta^2$ the assumed radar cross section. For the SP, the UE--SP leg contributes a delay $\|\boldsymbol{x}_{\mathrm{UE}}-\boldsymbol{x}_{\mathrm{SP}}\|/c$ that is identical at every AP; it is absorbed into $\theta^{m}$ and hence does not appear in $\tau^{m,n}$, whereas it does affect the gain. All APs are assumed to observe every modeled component.}

\begin{table}[t]
\centering
\caption{{Delay and gain of the three modeled component types, written in terms of the equivalent source $\boldsymbol{x}^{m}$ and $\dap{\boldsymbol{x}}=\|\boldsymbol{x}-\boldsymbol{x}^{n}_{\mathrm{AP}}\|$.}}
\label{tab:propagation}
\renewcommand{\arraystretch}{1}
\scriptsize
{\begin{tabular}{@{}l@{\hspace{2mm}}c@{\hspace{2mm}}c@{\hspace{1mm}}c@{}}
\hline
\textbf{Component} & $\boldsymbol{x}^{m}$ & $\tau^{m,n}$ & $(\rho^{m,n})^{2}$\\ \hline
LoS (UE) & $\boldsymbol{x}_{\mathrm{UE}}$ & $\dfrac{\dap{\boldsymbol{x}^{m}}}{c}$ & $\dfrac{\lambda^{2}}{(4\pi)^{2}\,\dap{\boldsymbol{x}^{m}}^{2}}$\\
Reflector (VUE) & $\boldsymbol{x}_{\mathrm{VUE}}$ & $\dfrac{\dap{\boldsymbol{x}^{m}}}{c}$ & $\dfrac{\lambda^{2}\Gamma^{2}}{(4\pi)^{2}\,\dap{\boldsymbol{x}^{m}}^{2}}$\\
Scatterer (SP) & $\boldsymbol{x}_{\mathrm{SP}}$ & $\dfrac{\dap{\boldsymbol{x}^{m}}}{c}$ & $\dfrac{\lambda^{2}\beta^{2}}{(4\pi)^{3}\,\dap{\boldsymbol{x}^{m}}^{2}\|\boldsymbol{x}_{\mathrm{UE}}-\boldsymbol{x}_{\mathrm{SP}}\|^{2}}$\\ \hline
\end{tabular}}
\end{table}

\section{Phase-Coherent Processing for SLAM} \label{Sec:phase}

\subsection{Principle of phase-coherent processing}
{The central unit stacks the observations as $\boldsymbol{y}=[y^1,\ldots,y^N]^{\mathsf T}$. Using the equivalent sources $\boldsymbol{x}^{m}$ of Table~\ref{tab:propagation}, the path model is}
\begin{align}
    {\boldsymbol{y}=\sum_{m=0}^{M}\boldsymbol{\gamma}^{m}\odot\boldsymbol{a}(\boldsymbol{x}^{m})+\boldsymbol{w},}\label{eq:signal_APs}
\end{align}
{The vector $\boldsymbol{\gamma}^{m}\in\mathbb{C}^N$ is an unknown complex path-amplitude vector with entries}
\begin{align}
    {[\boldsymbol{\gamma}^{m}]_n=\sqrt{E}\rho^{m,n}e^{\jmath\theta^m}\zeta,}\label{eq:allphi}
\end{align}
{and the location-dependent steering vector is}
\begin{align}
    {[\boldsymbol{a}(\boldsymbol{x})]_n=e^{-\jmath2\pi f_{\rmc}\|\boldsymbol{x}-\boldsymbol{x}^{n}_{\mathrm{AP}}\|/c}.}\label{eq:alla}
\end{align}
{For an SP, the UE--SP propagation phase is common to all APs and is included in $\theta^m$; only the SP--AP phase remains in $\boldsymbol{a}(\boldsymbol{x}^m)$. For a VUE, the AP--VUE distance already represents the complete reflected path. Thus, the relative phases across APs encode the hypothesized equivalent-source geometry. This is analogous to an array manifold, but the aperture is set by the geographic AP deployment.}

\subsection{Phase-coherent distributed positioning and mapping}
We form a grid-based coherent image $I_{\mathrm{MF}}:\tilde{\mathcal{X}}\rightarrow\mathbb{R}^{+}$ over candidate equivalent-source positions $\tilde{\boldsymbol{x}}\in\tilde{\mathcal{X}}$, with $\boldsymbol{a}(\tilde{\boldsymbol{x}})$ given by \eqref{eq:alla}, as
\begin{equation}
 {I_{\mathrm{MF}}(\tilde{\boldsymbol{x}})=\left|\boldsymbol{a}^{\mathrm H}(\tilde{\boldsymbol{x}})\boldsymbol{y}\right|^2,}\label{compute_correlation}
\end{equation}
which is a spatial matched filter, and it has an explicit likelihood interpretation under the single-source model $\boldsymbol{y}=\alpha\boldsymbol{a}(\boldsymbol{x})+\boldsymbol{w}$ with unknown complex $\alpha$ and $\boldsymbol{w}\sim\mathcal{CN}(\boldsymbol{0},N_0\boldsymbol{I})$.  Concentrating the log-likelihood over $\alpha$ gives $\hat{\alpha}=\boldsymbol{a}^{\mathrm H}(\tilde{\boldsymbol{x}})\boldsymbol{y}/N$ and leaves \eqref{compute_correlation} as the statistic to be maximized, since $\|\boldsymbol{a}(\tilde{\boldsymbol{x}})\|^2=N$ for all $\tilde{\boldsymbol{x}}$. Thus, \eqref{compute_correlation} is the \ac{ML} image for one source of AP-invariant gain. However, it is not the joint ML estimator for the $(M+1)$-component model of \eqref{eq:signal_APs}, where other paths are structured interference and the per-AP amplitude variation in $\boldsymbol{\gamma}^{m}$ is unexploited.

\subsection{Discussion of interference}
{When $\tilde{\boldsymbol{x}}=\boldsymbol{x}^m$, the phase of $\boldsymbol{a}(\tilde{\boldsymbol{x}})$ aligns with component $m$, producing a large value of $I_{\mathrm{MF}}$ when that path is sufficiently strong. A mismatch dephases the sum. Fig.~\ref{fig:example} illustrates this effect, but it also shows the important limitation: unequal path strengths let sidelobes from a strong component exceed the mainlobe of a weak component. Thus, a peak in the image is not by itself a reliable landmark detection.}
\begin{figure}%[htbp]
\center
\input{example2.tex}
\vspace{-1mm}
\caption{{One-dimensional illustrative example of $I(\tilde{x})$ with a UE and/or an SP, where the UE and SP are located at $0\,\text{m}$ and $4\,\text{m}$, respectively.}
}
\label{fig:example}
\end{figure}
{One useful consequence of the equivalent-source form is that a single steering vector tests UE, VUE, and SP hypotheses alike, so no separate estimator per landmark class is needed. A VUE is recovered as an equivalent source; converting it into the surface itself additionally requires the UE position, which the same image provides. For an SP, the UE--SP delay is AP invariant and does not affect the spatial correlation, so the SP--AP term alone produces the image peak. These properties follow from the idealized model of Sec.~\ref{sec:receivedsignal}.}

\subsection{Synchronization and resource implications}\label{sec:practical}
Let $\epsilon^n$ be the residual phase error at AP $n$ after synchronization and calibration, so that $y^n=\gamma_n e^{\jmath\epsilon^n}[\boldsymbol{a}(\boldsymbol{x})]_n+w^n$. Evaluating \eqref{compute_correlation} at a correct hypothesis gives, for an AP-invariant gain $\gamma_n=\gamma$, the noiseless value $|\gamma|^2\vert\sum_{n=1}^{N}e^{\jmath\epsilon^n}\vert^2$, against $N^2|\gamma|^2$ when $\boldsymbol{\epsilon}=\boldsymbol{0}$. A phase error common to all APs is therefore immaterial, whereas AP-dependent errors reduce the peak. Dividing by the error-free value $N^{2}|\gamma|^{2}$ expresses this as the retained fraction of coherent gain, $\eta_\phi^2(\boldsymbol{\epsilon})=\vert \frac{1}{N}\sum_{n=1}^{N}e^{\jmath\epsilon^n}\vert ^2\in[0,1]$, which equals one exactly when all $\epsilon^n$ coincide\footnote{Operationally, $\eta_\phi^2$ scales the mainlobe that the detector maximizes and compares against its threshold, so it acts as an effective SNR loss in the detection step. It likewise enlarges the cancellation residual, since the fit assumes a phase common to all APs.}.

{For independent $\epsilon^n\sim\mathcal{N}(0,\sigma_\phi^2)$ in this equal-gain model, $\mathbb{E}[\eta_\phi^2]=N^{-1}+(1-N^{-1})e^{-\sigma_\phi^2}$. Since this is decreasing in $N$ with limit $e^{-\sigma_\phi^2}$, the coherent loss is at most $10\log_{10}(e)\,\sigma_\phi^2\approx4.34\,\sigma_\phi^2\,$dB for \emph{any} $N$, with $\sigma_\phi$ in radians. It therefore stays below $1\,$dB for $\sigma_\phi\leq25^\circ$, and the degradation is graceful rather than abrupt: the array does not lose coherence at a threshold, it loses gain quadratically in the phase jitter. We set $\epsilon^n=0$ throughout, which is therefore an upper bound on the performance; a deployment additionally requires a synchronization and calibration architecture with periodic drift tracking, whose signaling and fronthaul cost we do not model.}

\section{Single-Snapshot Phase-Coherent SLAM} \label{Sec:phaseSLAM}

\subsection{Spatial ambiguity function and resolution}
{To characterize geometry-limited resolution independently of path power and noise, we use the normalized spatial ambiguity function \cite{Ambig1_98} between a hypothesized location $\tilde{\boldsymbol{x}}$ and a true equivalent-source position $\boldsymbol{x}$:}
\begin{align}
{\mathcal{A}(\tilde{\boldsymbol{x}},\boldsymbol{x})=
\frac{|\boldsymbol{a}^{\mathrm H}(\tilde{\boldsymbol{x}})\boldsymbol{a}(\boldsymbol{x})|}
{\|\boldsymbol{a}(\tilde{\boldsymbol{x}})\|\|\boldsymbol{a}(\boldsymbol{x})\|}.}\label{eq:AF}
\end{align}
It depends only on AP geometry, carrier frequency, and the propagation model. A narrow mainlobe improves spatial discriminability, while high sidelobes can mask weak paths or yield false peaks. Because $\boldsymbol{y}$ is a weighted sum of components, weak components may fall below strong-path sidelobes; selecting the largest $M+1$ peaks is therefore not a valid detector.

\subsection{Snapshot phase-coherent SLAM algorithm}
{The one-shot baseline is the spatial matched filter in \eqref{compute_correlation}. Its strongest peak can mask weaker components. We therefore use a greedy matching-pursuit-type detect-and-cancel procedure \cite{mallat1993matching,tropp2007signal}. 
Initialize}
\begin{equation}
{\boldsymbol{y}^{(0)}=\boldsymbol{y}.}
\end{equation}
{At iteration $t$, form $I^{(t-1)}(\tilde{\boldsymbol{x}})=|\boldsymbol{a}^{\mathrm H}(\tilde{\boldsymbol{x}})\boldsymbol{y}^{(t-1)}|^2$ and choose}
\begin{equation}
{\hat{\boldsymbol{x}}^{(t)}=\arg\max_{\tilde{\boldsymbol{x}}\in\tilde{\mathcal{X}}}I^{(t-1)}(\tilde{\boldsymbol{x}}).}\label{est_pos_music}
\end{equation}
{We consider two different cancellation methods that do not require knowledge of the calibrated amplitude profile. Incorporating calibrated-amplitude modeling would require additional information, such as the UE orientation and radiation pattern, and is therefore left for future work.} 
\subsubsection{\textit{Estimator I}} \rev{Once the $t$-th target location is detected, its contribution to the received signal is estimated and subsequently removed.  An intermediate complex path loss is first computed via
\begin{align}
\tilde{\boldsymbol{\gamma}}^{(t)} 
&= \boldsymbol{y}^{(t-1)}\odot \boldsymbol{a}^{*} (\hat{\boldsymbol{x}}), 
\end{align}
and the residual received signal is updated as
\begin{equation}
{\boldsymbol{y}}^{(t)} = {\boldsymbol{y}}^{(t-1)} - \hat{\boldsymbol{\gamma}}^{(t)}e^{j\hat\theta^{(t)}}\odot \boldsymbol{a}(\hat{\boldsymbol{x}}),\label{sub_signal}
\end{equation}
where $\hat\theta^{(t)}=\frac{1}{2}\angle\big(\sum_{n=1}^{N}(\tilde{\boldsymbol{\gamma}}^{(t)}_n)^{2}\big)$ and $\hat{\boldsymbol{\gamma}}^{(t)}=\Re\!\left\{\tilde{\boldsymbol{\gamma}}^{(t)}e^{-j\hat\theta^{(t)}}\right\}$ are the \ac{ML} estimates of the common phase shift and the real-valued path amplitude.}

\begin{remark}[\rev{Free-amplitude saturation of Estimator I}]\label{rem:saturation}
\rev{Estimator I fits $N$ per-AP amplitudes plus a common phase, consuming $N+1$ real degrees of freedom per component against only $2N$ available. Two components already saturate the model, so any two candidate positions fit the data exactly and the residual carries no positional information.}
\end{remark}

\subsubsection{\textit{Estimator II}} \rev{To address the free-amplitude saturation issue in Estimator I, we propose an alternative estimator with fewer degrees of freedom. In this formulation, the cancellation coefficient is computed as \begin{equation}\tilde{{\gamma}}^{(t)}=\frac{\boldsymbol{a}^{\text{H}}(\hat{\boldsymbol{x}}^{(t)})\boldsymbol{y}^{(t-1)}}{\boldsymbol{a}^{\text{H}}(\hat{\boldsymbol{x}}^{(t)})\boldsymbol{a}(\hat{\boldsymbol{x}}^{(t)})}=\frac{\boldsymbol{a}^{\text{H}}(\hat{\boldsymbol{x}}^{(t)})\boldsymbol{y}^{(t-1)}}{N}.\label{eq:coef}
\end{equation}
The residual update is then given by
\begin{equation}
{\boldsymbol{y}^{(t)}=\boldsymbol{y}^{(t-1)}-\tilde{{\gamma}}^{(t)}\boldsymbol{a}(\hat{\boldsymbol{x}}^{(t)}).}\label{sub_signal2}
\end{equation}
Therefore, Estimator II is a mismatched estimator that assumes AP amplitudes are flat, which is the model we use to interpret \eqref{compute_correlation}, while Estimator I leaves them free.}

\begin{remark}[Two mechanisms limit Estimator II]
The subtraction in \eqref{sub_signal2} is exact only if $\hat{\boldsymbol{x}}^{(t)}$ coincides
with the true equivalent source \emph{and} the component's gain is AP-invariant.
Suppose $\boldsymbol{y}^{(t-1)}$ contains a component $\boldsymbol{\gamma}\odot\boldsymbol{a}(\boldsymbol{x})$ with
$[\boldsymbol{\gamma}]_n=g^ne^{\jmath\theta}$, $g^n\ge0$, and that \eqref{est_pos_music} selects
the nearest grid point $\hat{\boldsymbol{x}}$. Writing $\varphi^n=2\pi f_c\Delta\tau^n$ with
\label{rem:grid}
\begin{equation}\label{eq:dtau}\Delta\tau^n=\frac{\|\hat{\boldsymbol{x}}-\boldsymbol{x}^n_{\mathrm{AP}}\|-\|\boldsymbol{x}-\boldsymbol{x}^n_{\mathrm{AP}}\|}{c},
\end{equation}
the cancellation coefficient \eqref{eq:coef} is
$\tilde\gamma=\frac{1}{N}\sum_n[\boldsymbol{\gamma}]_ne^{\jmath\varphi^n}$, and, because the
entries of $\boldsymbol{a}(\cdot)$ have unit modulus, the residual $\boldsymbol{e}$ left by
\eqref{sub_signal2} has normalized energy
\begin{equation}\label{eq:gridresidual}
  \frac{\|\boldsymbol{e}\|^2}{\|\boldsymbol{\gamma}\|^2}
  =1-\frac{\big|\sum_{n=1}^{N}g^ne^{\jmath\varphi^n}\big|^2}{N\sum_{n=1}^{N}(g^n)^2}\in[0,1),
\end{equation}
Two distinct effects appear in \eqref{eq:gridresidual}. \emph{(i) Grid mismatch\footnote{\rev{Estimator I is also affected by grid mismatch.}}:} For a flat profile ($g^{n}=g, \forall n$), \eqref{eq:gridresidual} becomes
$1-|N^{-1}\sum_n e^{\jmath\varphi^n}|^2$. Since $c\,\Delta\tau^n$ is a path-length mismatch, the $n$th phase error is $2\pi c\Delta\tau^n/\lambda$. A mismatch of $\lambda/6$  rotates that entry by $60^\circ$. \emph{(ii) Amplitude taper:} If $\hat{\boldsymbol{x}}$ is perfectly estimated,  $\varphi^n=0$, and \eqref{eq:gridresidual} becomes
$\sigma_{g}^{2}/(\bar{g}^{2}+\sigma_{g}^{2})$, with $\bar{g}$ and $\sigma_{g}^{2}$ being the mean and variance of $g^{n}$ across APs, which does not vanish as the grid is refined. Neither term is improved by raising the SNR. 

%\end{itemize}
The number of components need not be known: the recursion stops when $\max_{\tilde{\boldsymbol{x}}}I^{(t-1)}(\tilde{\boldsymbol{x}})<\kappa_{\mathrm{FA}}$, where $\kappa_{\mathrm{FA}}$ is calibrated on noise-only data for a desired false-alarm probability, or after $T_{\max}$ iterations. A controlled study may instead fix the iteration count to the number of modeled components, which isolates the mismatch effect from detector-threshold effects at the cost of not exercising the stopping rule. One image evaluation costs $\mathcal{O}(GN)$ complex operations for $G=|\tilde{\mathcal{X}}|$ grid points, so $T$ extracted components cost $\mathcal{O}(TGN)$, with $\mathcal{O}(N)$ working memory if steering vectors are generated on demand (or $\mathcal{O}(GN)$ if cached). The cost is thus linear in the number of APs and, being a grid search, embarrassingly parallel. 
\end{remark}

\vspace{-2mm}
\section{Results} \label{sec:results}

\subsection{Simulation environment}
{We use a controlled indoor scenario of $20\,\text{m}\times10\,\text{m}\times2.9\,\text{m}$, summarized in Table~\ref{tab:setup}. The APs are on the ceiling, and all target heights are fixed and known, so the estimator performs a two-dimensional search; the extension to a 3D search is direct but multiplies the grid size. The propagation model is deliberately sparse, with one LoS component, one visible planar reflector (represented by a VUE), and one SP, so that the mismatch mechanism of \eqref{eq:gridresidual} can be observed in isolation. Richer multipath from floor, walls and furniture, 3D ray tracing, and measured data are natural next validation steps.}
{Although the nominal allocation is $30\,$MHz, the imaging receiver retains and processes only one $30\,$kHz subcarrier, i.e., $0.1\%$ of the allocated bandwidth, leaving the remainder available for communication.  The $0.2\,$m ($2\lambda$) criterion in  Table~\ref{tab:setup} is a coarse acquisition criterion rather than a claim of wavelength-scale accuracy.}

\begin{table}[!b]
\centering
\caption{{Simulation setup.}}
\label{tab:setup}
\renewcommand{\arraystretch}{1.2}
\scriptsize
{\begin{tabular}{@{}p{0.30\linewidth}@{\hspace{1.5mm}}p{0.65\linewidth}@{}}
\hline
\textbf{Parameter} & \textbf{Value}\\ \hline
$f_{\rmc}$, $\lambda$ & $3\,$GHz, $0.1\,$m\\
Retained sensing band & $30\,$kHz (one subcarrier of a $30\,$MHz allocation)\\
Samples per AP & $1$ complex sample, $1$ antenna, $1$ snapshot\\
$N$, AP spacing & $50$, $2\,$m (ceiling, centralized collection)\\
Room & $20\times10\times2.9\,$m$^3$\\
Transmit power & $10\,$dBm\\
$N_0$, noise figure & $-174\,$dBm/Hz, $8\,$dB\\
%$\Gamma$, $\beta^2$ & $0.5$, $10\,$m$^2$\\
$M+1$ & $3$ (LoS, VUE, SP; single bounce, all visible, static)\\
Grid spacing ($\Delta$)& $1\,$cm (2D, known heights)\\
$\epsilon^n$ & $0$ (except Fig.~\ref{fig:offset})\\
{$\kappa_{\mathrm{FA}}$} & {$1\times10^{-5}$}\\
%$T$ & $3$ (known modeled component count)\\
Trials, success radius & $500$, $0.2\,$m ($2\lambda$)\\ \hline
\end{tabular}}
\end{table}

\vspace{-2mm}

\subsection{Results and discussion}
\begin{figure}%[htbp]
\center
\input{Detection_probability_dB.tex}
\vspace{-2mm}
\caption{Detection probability versus UE–VUE dynamic range, equal to the SP–VUE dynamic range. ``E.'' is Estimator. ``P.D.'' denotes peak detection, where the three largest local maxima of $I^{(0)}$ are selected without cancellation. }
\label{fig:detection_dB}
\end{figure}

{Fig.~\ref{fig:detection_dB} reports the detection probabilities of the two cancellation estimators versus the UE--VUE  dynamic range (set equal to the VUE--SP dynamic range), defined as the ratio of the total powers of the corresponding paths, i.e., 
$10\log10({\Vert\boldsymbol{\gamma}^{m_1} \Vert^2}/{\Vert \boldsymbol{\gamma}^{m_2}\Vert^2 })$. The UE is always detected since it forms the dominant peak. As the VUE and SP weaken, they become increasingly difficult to identify. Estimator~I fails to reliably recover the SP because of the free-amplitude saturation effect (Remark~\ref{rem:saturation}). In contrast, Estimator~II can detect both the VUE and the SP when the dynamic range is not large, but its performance is ultimately constrained by the two residual mechanisms in Remark~\ref{rem:grid}. Moreover, VUE detection with Estimator~I is generally stronger than with Estimator~II, since Estimator~II is affected by the amplitude taper across APs. As an additional baseline, selecting the three largest local maxima of the one-shot image (i.e., without cancellation) often misses the VUE and SP, since UE sidelobes can exceed the VUE mainlobe (and the SP is even weaker) except at very small dynamic ranges.

Fig.~\ref{fig:detection} reports the detection performance versus the search-grid spacing $\Delta$ over random trials. As expected, finer grids yield higher detection probabilities. The UE is selected before any cancellation, so its degradation with coarser grids is mainly due to undersampling. The grid can step over the narrow mainlobe and the sampled maximum may fall on a sidelobe. Weaker targets are more sensitive to this effect and, on coarse grids, also suffer from imperfect cancellation. This matches Remark~\ref{rem:grid}: for example, at $0.1\,$m spacing the path-length mismatch can reach several multiples of $\lambda/6$, so the cancellation residual can dominate the component being removed and the resulting error propagates to later iterations.}
The dashed oracle curves, labeled perfect removal (P.R.), isolate the cause. They subtract the true prior components and thus remove both terms of Remark~\ref{rem:grid} ($\boldsymbol{e}=\boldsymbol{0}$) while leaving the noise unchanged, and the SP then becomes nearly as detectable as the UE. At this noise level, the binding constraint is therefore the cancellation residual, not the noise floor, and the gap is a direct measure of error propagation. This indicates that better cancellation-error mechanisms would unlock weak-target mapping. In addition, the different detection-probability behavior of the VUE arises because the VUE lies outside the AP aperture and therefore behaves more like a far-field source.

{Fig.~\ref{fig:offset} illustrates the impact of residual per-AP phase jitter on UE detection for several grid resolutions. For all grid resolutions, the loss is smooth: increasing $\sigma_\phi$ gradually reduces the coherent combining gain rather than causing an abrupt failure. The decay is approximately quadratic, and the corresponding coherent loss remains limited for $\sigma_\phi\leq25^\circ$. With fine grids, detection is largely insensitive to moderate residual phase errors, whereas with coarser grids the performance is already grid-limited and the additional phase jitter translates into a more noticeable reduction in detection probability.}

\begin{figure}%[htbp]
\center
\input{Detection_probability_v2.tex}
\vspace{-2mm}
\caption{Detection probability versus search-grid spacing $\Delta$. ``P.R.'' denotes oracle perfect removal of previously detected components, i.e., $\boldsymbol{e}=\boldsymbol{0}$ in \eqref{eq:gridresidual}, and upper-bounds the achievable cancellation.}
\label{fig:detection}
\end{figure}

\begin{figure}%[htbp]
\center
\input{Detection_probability_phaseoffset.tex}
\vspace{-2mm}
\caption{{Detection probability for the UE versus $\sigma_\phi$ for different grid spacings.}
}
\label{fig:offset}
\end{figure}
\section{Conclusions}

In this paper, we studied uplink radio SLAM in a phase-coherent D-MIMO system in which each AP contributes only one complex sample, on one subcarrier, from one antenna. We formulated the problem as coherent matched-filter imaging over a grid of equivalent sources, so that the UE, planar reflectors and point scatterers are tested by a single steering-vector form, and we resolved them with a greedy detect-and-cancel receiver. Three conclusions follow. First, the results show that a single narrowband snapshot can suffice for joint localization and mapping under the considered model, where the geographic aperture of the AP network compensates for the absence of per-AP delay resolution. Second, phase coherence is demanding, but is shown not to be brittle. Third, the ability to recover weak targets is governed by the cancellation residual {and model saturation} rather than by the SNR.

\balance 
\bibliography{IEEEabrv,Bibliography}

\end{document}

%% file: acronyms.tex
\acrodef{MMSE}{Minimum Mean Squared Error}

\acrodef{MSE}{mean square error}

\acrodef{PSD}{power spectral density}

\acrodef{RMSE}{root mean squared error}
\acrodef{SLR}{statistical linear regression}

\acrodef{AP}[AP]{access point}

\acrodef{ML}[ML]{maximum likelihood}
\acrodef{DBSCAN}[DBSCAN]{density-based spatial clustering of applications with noise}
\acrodef{UE}[UE]{user equipment}
\acrodef{BS}[BS]{base station}
\acrodef{VA}[VA]{virtual anchor}
\acrodef{VUE}[VUE]{virtual user equipment}
\acrodef{SP}[SP]{scattering  point}
\acrodef{IP}[IP]{incidence point}
\acrodef{fov}[FoV]{field-of-view}   
\acrodef{LoS}[LoS]{line-of-sight}
\acrodef{NLoS}[NLoS]{non-line-of-sight}
\acrodef{PMBM}[PMBM]{Poisson  multi-Bernoulli  mixture}
\acrodef{PMB(M)}[PMB(M)]{Poisson  multi-Bernoulli  (mixture)}
\acrodef{PMB}[PMB]{Poisson  multi-Bernoulli}
\acrodef{RFS}[RFS]{random finite set}
\acrodef{PPP}[PPP]{Poisson point process}
\acrodef{MBM}[MBM]{multi-Bernoulli  mixture}
\acrodef{MB}[MB]{multi-Bernoulli}

\acrodef{ekf}[EKF]{extended Kalman filter}
\acrodef{PDF}[PDF]{probability density function}

\acrodef{ckf}[CKF]{cubature Kalman filter}
\acrodef{rbp}[RBP]{Rao-Blackwellized particle}
\acrodef{gospa}[GOSPA]{generalized optimal subpattern assignment}
\acrodef{SLAM}[SLAM]{simultaneous localization and mapping}

\acrodef{TOA}[TOA]{time of arrival}
\acrodef{AOA}[AOA]{angles of arrival}
\acrodef{AOD}[AOD]{angles of departure}

\acrodef{IF}[IF]{information filter}
\acrodef{EIF}[EIF]{extended information filter}
\acrodef{kf}[KF]{Kalman filter}
\acrodef{PMF}[PMF]{probability mass function}
\acrodef{MAP}[MAP]{maximum a posteriori estimation}

\acrodef{DA}[DA]{data association}
\acrodef{OFDM}[OFDM]{Orthogonal Frequency Division Multiplexing}

\acrodef{PCRB}[PCRB]{posterior Cram{\'e}r-Rao bound}

\acrodef{EM}[EM]{expectation-maximization}
\acrodef{FOV}[FOV]{field of view}
\acrodef{RB}[RB]{Rao-Blackwellized }

\acrodef{ISAC}[ISAC]{integrated sensing and communication}

\acrodef{EK}[EK]{extended Kalman}
\acrodef{EKF}[EKF]{extended Kalman filter}

\acrodef{LMB}[LMB]{labeled multi-Bernoulli}
\acrodef{GLMB}[$\delta$-GLMB]{$\delta$-generalized labeled multi-Bernoulli}
\acrodef{PHD}[PHD]{probability hypothesis density}

\acrodef{OID}[OID]{optimal importance density}

\acrodef{MTT}[MTT]{multi-target tracking}

\acrodef{MCMC}[MCMC]{Markov chain Monte Carlo}

\acrodef{MH}[MH]{Metropolis-Hastings}

\acrodef{NMI}[NMI]{normalized mutual information}

\acrodef{CRB}[CRB]{Cram{\'e}r-Rao bound}

\acrodef{PIM}[PIM]{posterior information matrix}

\acrodef{OID}[OID]{optimal importance density}

\acrodef{MAE}[MAE]{mean absoluate error} 

\acrodef{FIM}[FIM]{Fisher information matrix}
\acrodef{PIM}[PIM]{posterior information matrix}
\acrodef{PEB}[PEB]{position error bound}
\acrodef{LEB}[LEB]{landmark error bound}
\acrodef{HEB}[HEB]{heading error bound}
\acrodef{CEB}[CEB]{clock bias error bound}

\acrodef{AoA}[AoA]{angles of arrival}
\acrodef{AoD}[AoD]{angles of departure}

\acrodef{URA}[URA]{uniform rectangular array}

\acrodef{RTT}{round-trip-time}

\acrodef{LS}{least-squares}

\acrodef{CDF}{cumulative distribution function}

\acrodef{AWGN}{additive white Gaussian noise}

\acrodef{MUSIC}{multiple signal classification}

\acrodef{S-MUSIC}{sequential multiple signal classification}
\acrodef{OMP}{orthogonal matching pursuit}
\acrodef{SCM}{spatial covariance matrix}

\acrodef{MF}{matched filtering}

\acrodef{AIC}{akaike information criterion}
\acrodef{MDL}{minimum description length}

%% file: Detection_probability_dB.tex
\definecolor{mycolor4}{rgb}{0.92900,0.69400,0.12500}%
\begin{tikzpicture}[scale=0.8\linewidth/14cm]

\begin{axis}[%
width=6.028in,
height=2.009in,
at={(1.011in,2.014in)},
scale only axis,
xmin=3,
xmax=20,
xlabel style={font=\color{white!15!black},font=\Large},
xlabel={Dynamic range [dB]},
ymin=0,
ymax=1,
ylabel style={font=\color{white!15!black},font=\Large},
ylabel={Detection probability},
axis background/.style={fill=white},
xmajorgrids,
ymajorgrids,
legend style={legend cell align=left, align=left, draw=white!15!black,font=\Large,fill=white,
        fill opacity=0.6,
        draw opacity=1,
        text opacity=1},
scaled x ticks=false,
xticklabel style={
    /pgf/number format/fixed,
    /pgf/number format/precision=2
}
]
\addplot [color=black, mark=+, mark size=4pt, line width=2.0pt]
  table[row sep=crcr]{%
3	1\\
4	1\\
5	1\\
6	1\\
7	1\\
8	1\\
9	1\\
10	1\\
11	1\\
12	1\\
13	1\\
14	1\\
15	1\\
16	1\\
17	1\\
18	1\\
19	1\\
20	1\\
};
\addlegendentry{UE, E.I, E.II, P.D.}

\addplot [color=blue, dashed, line width=2.0pt]
  table[row sep=crcr]{%
3	0.79\\
4	0.77\\
5	0.75\\
6	0.74\\
7	0.73\\
8	0.72\\
9	0.71\\
10	0.71\\
11	0.66\\
12	0.64\\
13	0.63\\
14	0.52\\
15	0.46\\
16	0.4\\
17	0.39\\
18	0.35\\
19	0.31\\
20	0.22\\
};
\addlegendentry{VUE, E.I}

\addplot [color=blue, mark=o, mark size=4pt, line width=2.0pt]
  table[row sep=crcr]{%
3	0\\
4	0\\
5	0\\
6	0\\
7	0\\
8	0\\
9	0\\
10	0\\
11	0\\
12	0\\
13	0\\
14	0\\
15	0\\
16	0\\
17	0\\
18	0\\
19	0\\
20	0\\
};
\addlegendentry{SP, E.I}

\addplot [color=red, dashed, line width=2.0pt]
  table[row sep=crcr]{%
3	0.88\\
4	0.87\\
5	0.86\\
6	0.85\\
7	0.72\\
8	0.68\\
9	0.57\\
10	0.57\\
11	0.35\\
12	0.29\\
13	0.24\\
14	0.21\\
15	0.09\\
16	0.01\\
17	0\\
18	0\\
19	0\\
20	0\\
};
\addlegendentry{VUE, E.II}

\addplot [color=red,mark=o, mark size=4pt, line width=2.0pt]
  table[row sep=crcr]{%
3	0.55\\
4	0.37\\
5	0.14\\
6	0.1\\
7	0.01\\
8	0\\
9	0\\
10	0\\
11	0\\
12	0\\
13	0\\
14	0\\
15	0\\
16	0\\
17	0\\
18	0\\
19	0\\
20	0\\
};
\addlegendentry{SP, E.II}

\addplot [color=mycolor4, dashed, line width=2.0pt]
  table[row sep=crcr]{%
3	0.58\\
4	0.56\\
5	0.39\\
6	0.19\\
7	0.11\\
8	0.05\\
9	0.04\\
10	0.03\\
11	0.02\\
12	0\\
13	0\\
14	0\\
15	0\\
16	0\\
17	0\\
18	0\\
19	0\\
20	0\\
};
\addlegendentry{VUE, P.D.}

\addplot [color=mycolor4, mark=o, mark size=4pt, line width=2.0pt]
  table[row sep=crcr]{%
3	0.07\\
4	0.01\\
5	0\\
6	0\\
7	0\\
8	0\\
9	0\\
10	0\\
11	0\\
12	0\\
13	0\\
14	0\\
15	0\\
16	0\\
17	0\\
18	0\\
19	0\\
20	0\\
};
\addlegendentry{SP, P.D.}

\end{axis}
\end{tikzpicture}%

%% file: Detection_probability_v2.tex
\definecolor{mycolor4}{rgb}{0.92900,0.69400,0.12500}%
\begin{tikzpicture}[scale=0.8\linewidth/14cm]

\begin{axis}[%
width=6.028in,
height=2.009in,
at={(1.011in,2.014in)},
scale only axis,
xmin=0.01,
xmax=0.1,
xlabel style={font=\color{white!15!black},font=\Large},
xlabel={Search-grid spacing $\Delta$ [m]},
ymin=0,
ymax=1,
ylabel style={font=\color{white!15!black},font=\Large},
ylabel={Detection probability},
axis background/.style={fill=white},
xmajorgrids,
ymajorgrids,
legend style={legend cell align=left, align=left, draw=white!15!black,font=\Large,fill=white,
        fill opacity=0.6,
        draw opacity=1,
        text opacity=1},
scaled x ticks=false,
xticklabel style={
    /pgf/number format/fixed,
    /pgf/number format/precision=2
}
]
\addplot [color=black, mark=+, mark size=4pt, line width=2.0pt]
  table[row sep=crcr]{%
0.01	1\\
0.02	1\\
0.03	1\\
0.04	0.978\\
0.05	0.906\\
0.06	0.766\\
0.07	0.672\\
0.08	0.612\\
0.09	0.532\\
0.1	0.432\\
};
\addlegendentry{UE, E.I, E.II}

\addplot [color=blue,dashed, line width=2.0pt]
  table[row sep=crcr]{%
0.01	0.76\\
0.02	0.6\\
0.03	0.49\\
0.04	0.24\\
0.05	0.23\\
0.06	0.18\\
0.07	0.18\\
0.08	0.13\\
0.09	0.11\\
0.1	0.1\\
};
\addlegendentry{VUE, E.I}

\addplot [color=blue, mark=o, mark size=4pt, line width=2.0pt]
  table[row sep=crcr]{%
0.01	0\\
0.02	0\\
0.03	0\\
0.04	0\\
0.05	0\\
0.06	0\\
0.07	0\\
0.08	0\\
0.09	0\\
0.1	0\\
};
\addlegendentry{SP, E.I}

\addplot [color=red, dashed, line width=2.0pt]
  table[row sep=crcr]{%
0.01	0.86\\
0.02	0.86\\
0.03	0.8\\
0.04	0.73\\
0.05	0.69\\
0.06	0.45\\
0.07	0.35\\
0.08	0.23\\
0.09	0.27\\
0.1	0.17\\
};
\addlegendentry{VUE, E.II}

\addplot [color=red,mark=o, mark size=4pt,  line width=2.0pt]
  table[row sep=crcr]{%
0.01	0.37\\
0.02	0.22\\
0.03	0.19\\
0.04	0.17\\
0.05	0.07\\
0.06	0.05\\
0.07	0.02\\
0.08	0.01\\
0.09	0.01\\
0.1	0\\
};
\addlegendentry{SP, E.II}

\addplot [color=black, dashed, line width=2.0pt]
  table[row sep=crcr]{%
0.01	0.914\\
0.02	0.902\\
0.03	0.894\\
0.04	0.86\\
0.05	0.856\\
0.06	0.838\\
0.07	0.83\\
0.08	0.812\\
0.09	0.796\\
0.1	0.758\\
};
\addlegendentry{VUE, P.R.}

\addplot [color=black, mark=o, mark size=4pt,  line width=2.0pt]
  table[row sep=crcr]{%
0.01	1\\
0.02	1\\
0.03	1\\
0.04	0.998\\
0.05	0.932\\
0.06	0.816\\
0.07	0.698\\
0.08	0.608\\
0.09	0.566\\
0.1	0.446\\
};
\addlegendentry{SP, P.R.}

\end{axis}
\end{tikzpicture}%

%% file: Detection_probability_phaseoffset.tex
\definecolor{mycolor3}{rgb}{0.49400,0.18400,0.55600}
\definecolor{mycolor4}{rgb}{0.92900,0.69400,0.12500}%
\begin{tikzpicture}[scale=0.8\linewidth/14cm]

\begin{axis}[%
width=6.028in,
height=2.009in,
at={(1.011in,2.014in)},
scale only axis,
xmin=0,
xmax=50,
xlabel style={font=\color{white!15!black},font=\Large},
xlabel={Standard deviation of the residual phase error $\sigma_\phi$ [deg]},
ymin=0.5,
ymax=1,
ylabel style={font=\color{white!15!black},font=\Large},
ylabel={Detection probability for the UE},
axis background/.style={fill=white},
xmajorgrids,
ymajorgrids,
legend style={at={(0.25,0.7)},legend cell align=left, align=left, draw=white!15!black,font=\Large},
scaled x ticks=false,
xticklabel style={
    /pgf/number format/fixed,
    /pgf/number format/precision=2
}
]
\addplot [color=blue, line width=2.0pt]
  table[row sep=crcr]{%
0	1\\
5	1\\
10	0.996\\
15	0.992\\
20	0.988\\
25	0.988\\
30	0.98\\
35	0.98\\
40	0.964\\
45	0.968\\
50	0.876\\
};
\addlegendentry{$\Delta=0.01$~m}

\addplot [color=red, mark=+, mark size=4pt, line width=2.0pt]
  table[row sep=crcr]{%
0	0.996\\
5	0.996\\
10	0.996\\
15	0.994\\
20	0.988\\
25	0.988\\
30	0.984\\
35	0.982\\
40	0.976\\
45	0.958\\
50	0.874\\
};
\addlegendentry{$\Delta=0.02$~m}

\addplot [color=mycolor4, mark=o, mark size=4pt, line width=2.0pt]
  table[row sep=crcr]{%
0	0.994\\
5	0.992\\
10	0.992\\
15	0.994\\
20	0.988\\
25	0.986\\
30	0.982\\
35	0.978\\
40	0.976\\
45	0.904\\
50	0.792\\
};
\addlegendentry{$\Delta=0.03$~m}

\addplot [color=mycolor3, dashed, line width=2.0pt]
  table[row sep=crcr]{%
0	0.988\\
5	0.986\\
10	0.982\\
15	0.982\\
20	0.98\\
25	0.98\\
30	0.946\\
35	0.93\\
40	0.858\\
45	0.772\\
50	0.644\\
};
\addlegendentry{$\Delta=0.04$~m}

\addplot [color=black, dashdotted, line width=2.0pt]
  table[row sep=crcr]{%
0	0.912\\
5	0.898\\
10	0.88\\
15	0.878\\
20	0.872\\
25	0.836\\
30	0.824\\
35	0.784\\
40	0.722\\
45	0.602\\
50	0.518\\
};
\addlegendentry{$\Delta=0.05$~m}
\end{axis}
\end{tikzpicture}%